\documentclass[prl,twocolumn,aps,showpacs,floatfix,superscriptaddress]{revtex4}
\usepackage{amssymb} 
\usepackage{amsmath}{ 
\usepackage{graphicx}
\usepackage{epsfig} 
\usepackage{dcolumn}
\usepackage{textcomp,rotating}

\begin{document}

\title{Independent particle descriptions of tunneling from a many-body
perspective}

\author{Giorgos Fagas}
\affiliation{
Tyndall National Institute, Lee Maltings, Prospect Row, Cork, Ireland
}
\author{Paul Delaney}
\affiliation{
School of Mathematics and Physics, Queen's University Belfast,
Belfast BT7 1NN, Northern Ireland
}
\author{James C. Greer}
\affiliation{
Tyndall National Institute, Lee Maltings, Prospect Row, Cork, Ireland
}

\date{\today}

\begin{abstract}
Currents across thin insulators are commonly taken as single electrons moving
across classically forbidden regions; this independent particle picture is
well-known to describe most tunneling phenomena. Examining quantum transport
from a different perspective, i.e., by explicit treatment of electron-electron
interactions, we evaluate different single particle approximations
with specific application to tunneling in metal-molecule-metal junctions.
We find maximizing the overlap of a Slater determinant composed
of single particle states to the many-body current-carrying state is 
more important than energy minimization for defining single particle
approximations in a system with open boundary conditions.
Thus the most suitable single particle effective potential is not one commonly
in use by electronic structure methods, such as the Hartree-Fock or Kohn-Sham
approximations.
\end{abstract}


\maketitle
Describing quantum transport from first-principles has proven to be a challenging
task; debate continues as to the proper theoretical
approach for treating electron currents across metal-molecule-metal junctions
(MMJs)~\cite{PRB06KBE,PRB04EWK,PRL05TFS,CFR05,PRL04DG,PRB05ZDD,con05ASS}.
Tunneling is an archetype of quantum behavior with implications for all branches 
of modern physics, as well as chemical and biological processes dominated by electron 
transfer. The ability to accurately describe tunneling currents on the nanoscale is
important for both fundamental science and technology perspectives.
The discussion focusses  on the suitability of various Kohn-Sham density
functional theory implementations
(KS-DFT)~\cite{PRB06KBE,PRB04EWK,PRL05TFS,CFR05},
or possibly Hartree-Fock method~\cite{PRB04EWK,PRB05ZDD},
to apply in combination with the one-body non-equilibrium Green's function (NEGF)
approach~\cite{CFR05,CP02XDR}. In fact though, there is no established criterion for 
selecting a single particle Hamiltonian to be used in transport calculations.

On the other hand, evidence has been mounting that electron transport can be 
sensitive to many-body exchange and correlation~\cite{PRB04EWK,PRB05ZDD,con05ASS}.
For example, applying NEGF together with static KS-DFT leads to overestimation
of the current in MMJs, in some cases, up to orders of magnitude when compared to
experiments~\cite{CFR05}. Several reasons for discrepancies have been suggested:
DFT's underestimation of occupied/unoccupied state separation
results in too high currents~\cite{PRL05TFS,CPL04FKE}, conversely
the over-estimation of the gap in the HF approximation yields currents
too low~\cite{PRB04EWK}; DFT exchange-correlation functionals do not accurately reflect
potential profiles~\cite{PRL04KKP}; other methods such as
time-dependent current DFT~\cite{PRL05SZV} or density matrix functional
theory are simply better suited for the treatment of electronic currents.
Irrespective of the proposals to explain/remedy single particle tunneling
descriptions, it is necessary to understand the physics
deriving from a genuine many-body formalism to identify the source of
the discrepancies.

In this Rapid Communication we investigate the extent of correlations beyond the
single particle picture and  identify conditions for defining a ``best'' independent
particle model, using a recently formulated many-body quantum transport
approach~\cite{PRL04DG}. We demonstrate a single electron description of tunneling
deriving from maximizing the overlap of a single Slater determinant
with the true many-body current carrying state, and show that it remains a
good approximation even after conventional independent particle models fail.

Most theoretical studies of quantum transport begin with the use of a Slater
determinant of single particle states to model a many-body current-carrying
wavefunction~\cite{PRL61B}. Our recent work~\cite{PRL04DG} finds this picture
becomes markedly less valid near or above a resonance. Here, we concentrate
on non-resonant tunnel models. Independent particle models are appealing
because as the wavefunction can be represented as a single Slater determinant,
the resulting physical model can readily be pictured and
computed~\cite{CFR05,JCP05ZPP}.
Their limitation is neglect of electron correlations; below we introduce a correlation
measure to quantify when an independent particle model holds.

\begin{figure}[t]
\includegraphics[width=0.875\linewidth, height=4.5cm]{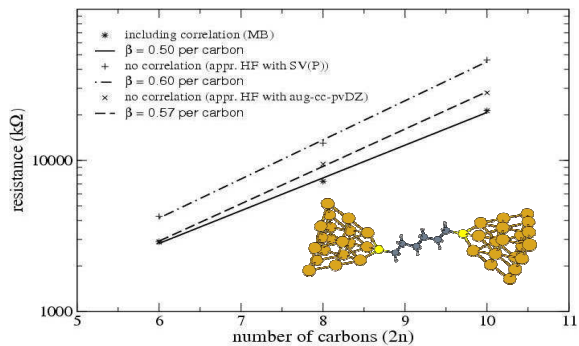}
\caption{
Tunnel resistance-increase exponential law. Explanation of the
standard basis set notations is given in the text.
Inset: Typical geometry of the studied
Au$_{tip}$-S-(C$_2$H$_4$)$_n$-S-Au$_{tip}$ molecular junctions ($n=3$).
}
\label{fig1}
\end{figure}

For our study, MMJs are an ideal test case as the molecule acts like an
insulator when there is little charge transfer or hybridization with the electrodes, 
there is a large set of experimental observations to validate the
calculations, and MMJs may be modelled with a relatively small number of atoms
allowing their electronic structure to be accurately calculated with many-body methods.
We choose alkane chains (C$_2$H$_4$)$_n$  (inset of Fig.~\ref{fig1}) 
and for comparative purposes also examine silicon hydride molecules
(Si$_2$H$_4$)$_n$ (silanes), as replacing the carbon atoms with silicon reduces
the occupied/unoccupied state separation and increases the degree of correlation.

Details of the transport method are given in ref.~\onlinecite{PRL04DG}.
A constraint {\it Ansatz} using the Wigner function is made to incorporate 
open-system boundary conditions for calculation of the
reduced density matrix on a device region (typically the molecule
plus part of the electrodes) at several values of applied voltage.
The procedure results in the best approximation to the
density matrix on a region subject to reproducing known system observables
in accord with the principle of maximum entropy at the zero temperature limit.
The density matrix is calculated from eigenfunctions of a constrained
many-electron Hamiltonian. A  strength of our approach is that it allows
for the expansion of many-body states in terms of a complete set of configurations
\begin{eqnarray}
| \Psi \rangle & =&  c_0 | \Psi_0\rangle +
         \sum_{i,a} c_i^a |\Psi_i^a\rangle +
\sum_{i<j}\sum_{a<b} c_{ij}^{ab} |\Psi_{ij}^{ab}\rangle
                  + \ldots
\label{Eq1}
\end{eqnarray}
$| \Psi_0\rangle$ refers to a reference state composed of $N$ lowest single
particle states, $|\Psi_i^a\rangle$ ($|\Psi_{ij}^{ab}\rangle \ldots$) denotes
singly- (doubly- $\ldots$) excited configurations generated by substituting the
$i$-th ($j$-th$, \ldots$) occupied single particle state with the $a$-th
($b$-th$, \ldots$) single particle excitation. Indexed $| \Psi \rangle $ are
spin-projected Slater determinants or configuration state functions (CSFs),
and $| \Psi_0\rangle$ is the ground-state HF determinant in the absence of an
applied electric field. The symmetry group of our contact-molecule-contact 
subsystem is $C_{2h}$  (inset of Fig.~\ref{fig1})
whose ground state eigenfunction at $V=0$ is a singlet with $A_g$ symmetry.
The bias field polarizes the molecular subsystem and mixes states of 
singlet $B_u$ symmetry via dipole terms.
Due to the combinatorial nature of generating possible configurations,
there is a computationally prohibitive number of excited state determinants.
To calculate the many-body tunneling wavefunction $|\Psi_{\rm MB}\rangle$
we approximate the full expansion~(\ref{Eq1}) by taking the most
significant CSFs,
as computed from a Monte Carlo sampling of the many-electron expansion
space~\cite{JCP98G} with the addition of all singly excited CSFs.
Notably, this approximation is not excitation limited and allows for a high 
degree of electron correlation~\cite{comment1}.

In contrast to the many-body wavefunction, a single Slater determinant 
$|\Psi_{1-{\rm det}}\rangle$ is sought in independent particle models.
If this is initially built from a set of {\it unperturbed} single particle states,
e.g., the Hartree-Fock $| \Psi_{0}\rangle$ as above, and a perturbation such as an
electric field is introduced, the resulting many-body expansion of the perturbed single
determinant state has the special feature that the coefficients of the doubly-(triply-,...) 
excited configurations will be appropriate products of the coefficients of the 
singly-excited configurations
\begin{eqnarray}
| \Psi_{1-{\rm det}}\rangle & \approx
&  | \Psi_{0}\rangle + \sum_{i,a} c_i^a |\Psi_i^a\rangle + 
\sum_{\substack{i < j\\a \not= b}} c_i^a c_j^b |\Psi_{ij}^{ab}\rangle  + \ldots \label{Eq2}
\end{eqnarray}
That this is the most general determinant not orthogonal
to $| \Psi_0\rangle$ is seen by expanding the Thouless expression 
$| \Psi_{1-{\rm det}}\rangle = {\mbox{exp}}(\sum_{i,a} c_i^a \hat a_a^\dagger \hat a_i )|\Psi_0\rangle $,
where $a_i^\dagger (a_i)$ creates (annihilates) the $i$-th orbital.
From the many-body calculation we find that there is a broad voltage range 
where linear response is valid; then by taking terms up to second order in the 
$c_i^a$ in~(\ref{Eq2}) and applying the same Wigner function transport scheme to 
this restricted form of the wavefunction, we deduce the HF results to linear response 
in the bias. This permits comparison of the exact-exchange (but uncorrelated) 
and correlated descriptions of tunneling within the same formalism.

In Fig.~\ref{fig1} both correlated and uncorrelated tunnel resistance 
of the alkane-based molecular junctions are shown. The qualitative features are
the exponential suppression $R=R_0 e^{\beta N_{\rm c}}$ of the current
with respect to the number of carbon atoms $N_{\rm c}=2n$ (Fig.~\ref{fig1})
and linear scaling with applied voltage. 
As observed in numerous experiments, both results are typical of a
metal-insulator-metal system. Contact resistances
$R_0$ and tunneling parameters $\beta$ calculated by least-square 
linear fits are relatively close in value with HF $R_0$ and $\beta$ deviating
from the many-body results by 32\% and 14\%, respectively.
The overestimation of $\beta$ by the HF approximation can be interpreted in terms of
the incorrect alignment of the virtual orbitals which leads to a larger highest
occupied-lowest unoccupied molecular orbital gap. In contrast,
the contact resistance $R_0$ is underestimated as 
indicated by the zero crossing in Fig.~\ref{fig1}, suggesting a stronger
molecule-electrode coupling within HF.
Similar behavior is found for the silane chains, with $\beta=0.18$ per silicon
and a contact resistance of $R_0=900$K$\Omega$ from the many-body calculation.
As expected, these molecular junctions are much more conducting yielding a value of
$\beta \approx 0.09$\AA$^{-1}$ compared to $\beta \approx 0.39$\AA$^{-1}$
for alkanes. But, as will be discussed in detail, the silanes display a higher degree
of electron correlation and the HF deviations with respect to the many-body results
are substantially larger than for the alkanes, increasing to 75\% and 117\% for
$R_0$ and $\beta$, respectively.

Quantitatively, the magnitudes of the low-bias resistance for each alkane molecule
compares well with those reported in ref.~\onlinecite{Sci03XT} apart from decane
which appears to be experimentally poorly resolved, as the measured result
for $n=5$ does not fit the extrapolated tunneling behaviour.
Our estimated inverse decay length
$\beta$ falls in the lower part of the experimentally observed range of values scattered 
between around 0.5 and 1.0 per carbon atom~\cite{AM03SCL}.
In particular, Haiss {\it et al.}~\cite{PCCP04HNZ} have reported
$\beta = 0.52 \pm 0.05$ per carbon but with higher resistance values (compare
to values indicated in Fig.~\ref{fig1}).

\begin{figure}[t]
\includegraphics[width=0.875\linewidth, height=4.5cm]{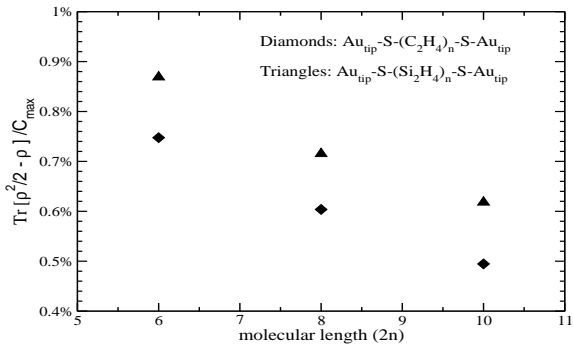}
\caption{
Many-body correlations expressed via the one-body density matrix.
These are voltage-independent with a standard
deviation much smaller than the symbol size.
}
\label{fig2}
\end{figure}
We can state that HF performs relatively well compared to a full many-body
calculation for non-resonant transport in the alkanes, but eventually
diverges from the many-body result as the molecular length increases.
Due to the higher degree of correlation, HF is a poorer approximation for the
silanes, and also diverges with chain length as compared to the many-body
calculation. In fig.~\ref{fig1}, the HF results for the smaller but commonly
used atomic orbital
split valence with polarization (SV(P)) basis set for the carbon atoms is included
(the larger aug-cc-pvDZ denotes a valence double zeta correlation-consistent basis
set with polarization, augmented with diffuse functions). Within this approximation,
the molecular gap becomes larger yielding a small increase in $\beta$ and
higher resistance. This result does not compare as well to the many-body calculation 
and casts doubts on results whose convergence has not been tested with respect to the
completeness of the single particle basis. 

We have examined the importance of electron correlation by comparing
separate calculations at the many-body and HF levels, however, the level of
correlation may be directly quantified from the many-body calculation
{\it independently} of HF results by examining the one-body density matrix
${\mathbf \gamma}_1$. Lack of correlations is expressed by the condition that
${\mathbf \gamma}_1$ may be derived from a single determinant if and only if
it is idempotent, ${\mbox Tr}[ \hat{\gamma}_1^2 - \hat{\gamma}_1 ] = 0.$
For closed shell systems such as studied here,
the degree of correlation $C \ge 0$ -- or equivalently the deviation from idempotency  --
for the {\it spin-traced} density matrix ${\mathbf \rho}_1$ may be defined
in terms of its eigenvalues or natural occupation numbers $n_i$ as
\begin{equation}
C =\frac{1}{N-1} \sum_i n_i (2- n_i)/2.
\end{equation}
As $0\le n_i \le 2$, an upper bound  can be deduced:
\begin{equation}
C_{max} = \frac{N}{2(N-1)} \left(2-\frac{N}{N_T}\right).
\end{equation}
Here $N$ is the total number of electrons in the device region and
$N_T$ is the size of the single particle basis.
The magnitude of $C$ is a direct measure of the deviation of the actual one-body
density matrix from an uncorrelated one deriving from a determinantal
wavefunction; $C = 0$ for a single determinant and $C=C_{max}$ when $n_i = N/N_T$.

\begin{figure}[t]
\includegraphics[width=0.875\linewidth, height=4.5cm]{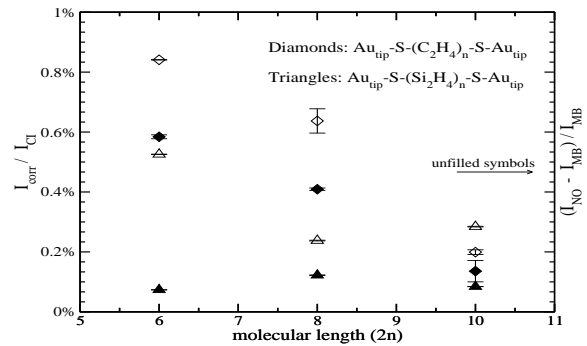}
\caption{
(Left axis) Contribution of correlations to the current measured via Eq.~\ref{Eq4}.
(Right axis) Current percentage {\it not} carried by the leading
determinant defined as the $N$-particle configuration with highest
natural occupancies. Residual noise voltage-dependence
is indicated by error bars.
}
\label{fig3}
\end{figure}

In fig.~\ref{fig2}, we plot the correlation measure $C$ scaled to its upper limit 
as a function of chain length for $| \Psi_{\rm MB} \rangle$ (note that
it vanishes for $| \Psi_{1-{\rm det}} \rangle$). 
Remarkably, as a function of voltage (not shown) the correlation measure
is constant and close to the uncorrelated limit for any given oligomer.
In fact, $C$ monotonically reduces with increasing length possibly approaching
a lower limit set by an infinite polymer for both the alkanes and
the silanes; the latter system displaying more electronic correlation.
Hence, in contrast to the HF predictions, the density matrix reveals
there exists a single Slater determinant description that is equally
valid for {\it all} chain lengths.

A formal measure of correlation for any expectation value
$\langle \Psi_{\rm MB} | \hat{I} | \Psi_{\rm MB} \rangle$ can be devised by using the
fact that the spin-traced two-body density matrix ${\mathbf \rho}_2$ 
factorizes in terms of ${\mathbf \rho}_1$ only in the one-determinant (uncorrelated)
approximation. We write
${\mathbf \rho}_2(ij;kl) = (1/2)({\mathbf \rho}_1(ik) {\mathbf \rho}_1(jl)-
{\mathbf \rho}_1(il){\mathbf \rho}_1(jk)/2) + \lambda(ij;kl)$,
where the first term is familiar from HF theory
and ${\bf \lambda}$ denotes any deviations when ${\mathbf  \rho}_1$ is not idempotent.
Here $i$ and $k$ are indices of the first electron, $j$ and $l$ that of the second and
$\sum_{i,j} {\mathbf \rho}_2(ij;ij) = N(N-1)/2$.
Tracing over the coordinates of the second electron, we deduce a partition of 
${\mathbf \rho}_1= 1/(N-1) [ N {\mathbf \rho}_1  - {\mathbf \rho}_1 \times {\mathbf \rho}_1/2 ] + \lambda$
into uncorrelated and correlated pieces. This gives us a decomposition of the current
$\langle{\hat I}\rangle = \mbox{Tr} [{\hat {\mathbf \rho}}_1 \hat{I}]$
flowing in the many-body wavefunction as $I_{\rm MB} = I_{1-{\rm det}} + I_{corr}$,
where $I_{corr}$ includes the contributions from ${\mathbf \lambda}$
and does not vanish when ${\mathbf \rho_1}$ derives from a correlated wavefunction.
It follows
\begin{equation}
I_{corr} =
\frac{1}{N-1} \mbox{Tr} [(\hat{\rho}_1^2/2 - \hat{\rho}_1)\hat I].
\label{Eq4}
\end{equation}
The percentage of current carried by the correlation terms is indicated in
fig.~\ref{fig3} by the filled symbols. In line with our previous observations,
it is characterised by a single value for each oligomer. The correlation
contributions remains a small fraction of the total current as a function of
applied voltage. These findings imply the existence of an approximate 
single-determinantal current-carrying wavefunction that captures most of the
contributions to electron flow even when the HF approximation begins to fail.

\begin{figure}[t]
\includegraphics[width=0.875\linewidth, height=5cm]{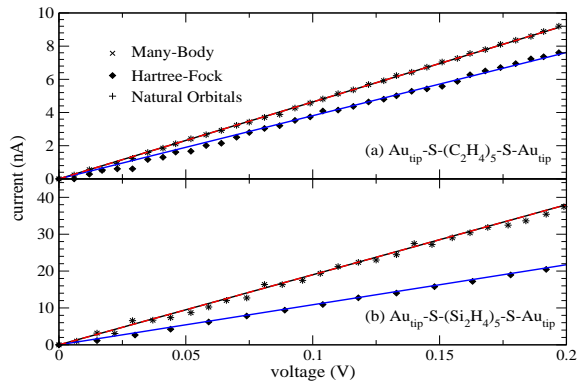}
\caption{
Comparison of the (a) Many-Body, (b) Hartree-Fock, and (c) leading
natural orbital determinant current.
}
\label{fig4}
\end{figure}

The HF approximation determines the best orbitals for a $|\Psi_{1-{\rm det}}\rangle$
that minimizes the energy. Since the current largely reflects the form of the
wavefunction, it is plausible that an approximate state that maximizes overlap 
with the current carrying wavefunction may improve
the orbital description of an independent particle model. Truncated configuration
expansions built from the eigenfunctions of the one-body density matrix, 
the natural orbitals (NOs), are known to maximize overlap to
$|\Psi_{MB}\rangle$. Since there are only $N/2$ NOs with large occupation numbers
$n_i$ --- in agreement with the calculated small correlation measure $C$ in our MMJs ---
we use the single determinant made by doubly occupying the orbitals with occupation
numbers close to 2; note that
an exact treatment of maximizing the overlap for $|\Psi_{1-{\rm det}}\rangle$
yields the Brueckner determinant. We find that the current carried
$I_{\rm NO}$ is close to $I_{\rm MB}$ as shown in fig.~\ref{fig3} where the
percentage of the remaining differences is plotted versus molecular length.
Remarkably, these deviations are of the same order as the correlation contributions 
determined by eq.~\ref{Eq4}. Indeed, in fig.~\ref{fig4} the current-voltage 
characteristics clearly demonstrate that the ``maximum overlap model'' yields almost
indistinguishable results from the many-body calculations independent
of residual electron correlations; the HF single-electron picture already 
shows significant deviation for pentane (fig.~\ref{fig4}(a))
which grows much larger for the silanes
(fig.~\ref{fig4}(b)). The use of the natural orbitals in defining
$|\Psi_{1-{\rm det}}\rangle$ acquires a practical significance in view of recent
advances allowing their construction from one-electron equations with an
effective potential~\cite{PRL05P}.

In summary, we have shown within a many-body scheme how an independent particle 
description emerges for electrons tunneling across the barrier of insulating
materials, due to zero-bias correlations being such that a single
determinant wavefunction is appropriate. Our main conclusion is that
a strong selection criterion for a single particle transport model
is to maximize the overlap with the many-body state;
this does {\it not} yield the Hartree-Fock determinant.
Indeed, this Slater determinant outperforms HF substantially.
As most KS-DFT implementations to date have recently been shown to be
performed at essentially the Hartree level~\cite{PRB06KBE} the independent
particle picture we have established here is better
than {\it both} present models and, hence, valid beyond the limits of
conventional electronic structure methods for quantum transport.

\begin{acknowledgments}
We would like to thank T.M. Henderson for many helpful discussions.
Technical assistance by A. Uskova and S.D. Elliott is acknowledged.
This work was funded by Science Foundation Ireland.
\end{acknowledgments}



\begin{references}
\bibitem{comment1}
Convergence of the results has been checked against the increase of the
CI-expansion. For the reported calculations, the number of CSFs is around
$10^4$.
\end{references}
\end{document}